\documentclass[epj]{svjour}
\usepackage{graphics}
\begin{document}
\title{The Detector and Interaction Region for a Photon Collider at TESLA}
\author{Aura Rosca
}                     
\institute{DESY Zeuthen, Platanenallee 6, 15738 Zeuthen, Germany}
%
\date{Received: October 3, 2003}
%
\abstract{
TESLA is designed as an electron-positron linear collider (LC) based on 
super-conducting technology. A second interaction region is forseen to be
incorporated in the design allowing its possible operation as a photon collider.
In this paper I describe the basic design of the $\gamma \gamma$ interaction 
region taking into account the beam-beam and laser related issuses and review 
some aspects of other accelerator components such as the feedback system and the beam 
dump which are critical to the operation of TESLA as a photon collider.  
} 
\maketitle
\section{Introduction}
\label{intro}

The collisions of high energy photon beams will provide new opportunities for
particle physics, ranging from particle searches to the measurement of the Higgs boson
properties. 

The cross sections for charged pair particle production in $\gamma \gamma$ collisions
are about one order of magnitude higher than in $\rm e^{+} \rm e^{-}$ collisions
\cite{cs}. 
Therefore, one can study
new particles far from threshold with higher rate. For example, WW pair production
at 500 GeV is a factor of 10 larger than in $\rm e^{+} \rm e^{-}$. About
2$\cdot 10^{6}$ WW pairs could be produced at a photon collider with a luminosity of
100 fb$^{-1}$ thus allowing a precise study of the anomalous gauge boson interactions.
Also, the cross sections for the production of charged scalar, lepton and top pairs are 
a factor of 5 to 10 higher at a photon collider than at an $\rm e^{+} \rm e^{-}$ LC
\cite{cs}.

But the main motivation for the construction of a photon collider is to
measure the properties of the Higgs bosons. A photon collider permits a direct measurement
of the two photon decay width of the Higgs boson with a precision better than 2$\%$
\cite{higgs} for a Higgs boson with a mass of 120 GeV.
The coupling of the Higgs to two photons proceeds through loops to which any charged
particle that gets its mass through the Higgs mechanism will contribute, no matter how
heavy it is. A measurement of the
$\gamma \gamma$ width is thus very sensitive to new physics. At a photon collider the
Higgs bosons are produced on resonance by the fusion of two photons.
This means a higher discovery reach for the heavier Higgs bosons. Very important is
the determination of the CP nature of the Higgs bosons which could be done at a photon
collider using linearly polarized photon beams.

\begin{figure}
\resizebox{0.3\textwidth}{!}{%
  \includegraphics{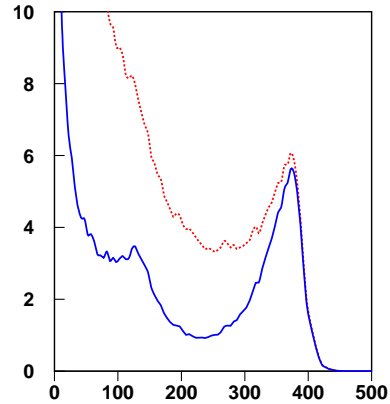}
}
\caption{$\gamma \gamma$ luminosity spectra at TESLA. The solid line shows the
luminosity distribution for total helicity of the photons 0 and the dotted line
for the case the total helicity 2 is also included.}
\label{fig:2}       
\end{figure}

The beams for a photon collider can be obtained by converting the high energy electron
beams at a LC into high energy photons through Compton
backscattering \cite{pc} of high density laser photons with an energy of approximative 
1 eV, a 
few millimeter before the interaction point. The high energy photons will follow the
direction of motion of the electrons and will collide at the interaction point.

The maximum energy of the generated photons is given by E$_{\gamma}^{\rm max}=
xE_{\rm e}/(1+x)$, with $E_{\rm e}$ being the electron beam energy and
$x=\frac{4 \omega_{\rm L} E_{\rm e}}{m_{\rm e}^{2}c^{4}}$, where $\omega_{\rm L}$,
$m_{\rm e}$ and $c$ are the laser photon energy, the electron rest mass, and 
the velocity of light, respectively. It can reach up to 80$\%$ of the initial electron 
beam energy.

The luminosity distribution in $\gamma \gamma$ collisions has a high energy peak,
see Figure \ref{fig:2}, which is the most useful for the measurements, and a low 
energy part.
The luminosity in the hard part of the spectrum can be up to 10$\%$ \cite{tdr} 
of the geometric
luminosity of the electron beam. 

The Compton scattered electrons will have a large energy spread and will undergo
a significant disruption during the beam-beam interaction. To ensure a clean
extraction of these electrons, the photon collider requires a crossing angle at the
interaction point, to allow a larger pipe for the outgoing beam. The current TESLA 
design has a crab crossing angle at the $\gamma \gamma$ interaction point of 34 mrad
\cite{tdr}.

To accomodate the extraction lines with larger apertures, as well as the laser beams in
the interaction region, the design of this region for $\rm e^{+} \rm e^{-}$ collisions
must be modified for a $\gamma \gamma$ collider. 
In addition, also a mask system which protects the vertex detector from the background 
caused by beam-beam interactions should be installed here.

In the remaining of this paper several technical issues of the photon collider will be
discussed, together with some open items which still require studies before the
construction of such a collider becomes feasible.

\section{The laser and optical systems}
\label{laser}

The laser system must match the bunch structure of the electron beam, which is 2800
bunches in a train with 5 trains per second
for TESLA, and to provide the required laser photon density at the
conversion point in order to convert efficiently the electrons into high energy 
photons. For efficient conversion of 250 GeV electrons, the optimal laser wavelength
is about 1 $\mu$m and the pulse energy is 5 J, with a pulse duration of about 1 ps.
The laser requirements for the TESLA photon collider are summarized in Table \ref{tab:1}.

\begin{figure}
\resizebox{0.5\textwidth}{!}{%
  \includegraphics{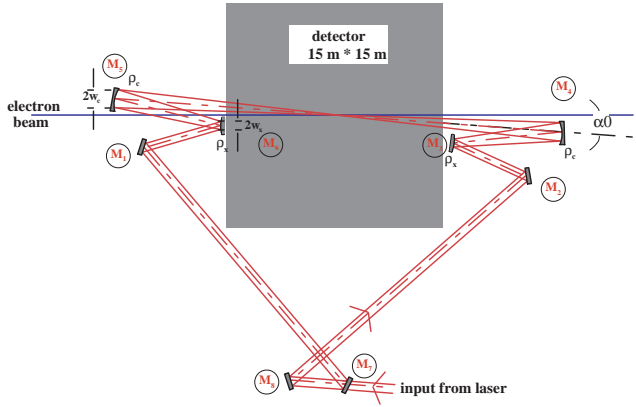}
}
\caption{A schematic view of the laser cavity proposed for the TESLA photon collider.}
\label{fig:1}       
\end{figure}

\begin{table}
\caption{The laser requirements}
\label{tab:1}       
\begin{tabular}{ll}
\hline\noalign{\smallskip}
Parameter & Requirement  \\
\noalign{\smallskip}\hline\noalign{\smallskip}
Laser wavelength & $\lambda \approx 1 \mu \rm m$  \\
Laser energy & $E_{\rm pulse} \approx$ 5 J  \\
Pulse duration & $\tau \approx$ 1 - 3 ps  \\
Rayleigh length & $Z_{\rm r} \approx$ 0.4 mm  \\
Repetition rate & TESLA collision rate  \\
Average power & $P \approx$ 70 kW  \\
\noalign{\smallskip}\hline
\end{tabular}
\end{table}

Due to the long pulse separation and the large number of pulses per train at TESLA,
the required average power of the laser system is very high and cannot be generated
at present. It has been proposed to increase the laser power by means of an optical
resonant cavity \cite{cavity} with a quality factor reaching 100. The length of the 
cavity
has to be such that the round trip of the laser pulses inside matches the bunch
separation of the electron bunches in the train, and has to be stabilized within 0.5 nm. 
This results in about 100 m cavity length at TESLA. Figure \ref{fig:1} shows a drawing of the optical
cavity consisting of several mirrors, all mounted around the detector. 
The laser radiation is transfered into the cavity through
a coupling mirror and brought to the interaction region with two
final focusing mirrors placed away from the beam pipe. 
The crossing angle between the laser pulse and the electron beam
is determined by how small the focusing mirrors can be realised without producing
significant losses of the laser pulses circulating inside the cavity. The laser
collision angle reduces the conversion probability which can be compensated by a higher
laser energy. A value of about 60 mrad \cite{angle} is forseen in the present design.

Laser pulses with an energy of several Joules and a duration of the order of picoseconds
can be achieved with modern laser technology. The wavelength of the most powerful
solid-state lasers is about 1 $\mu$m, which is exactly the required value for the TESLA
photon collider.

\section{Backgrounds and the detector design}
\label{detector}

\begin{figure}
\resizebox{0.45\textwidth}{!}{%
  \includegraphics{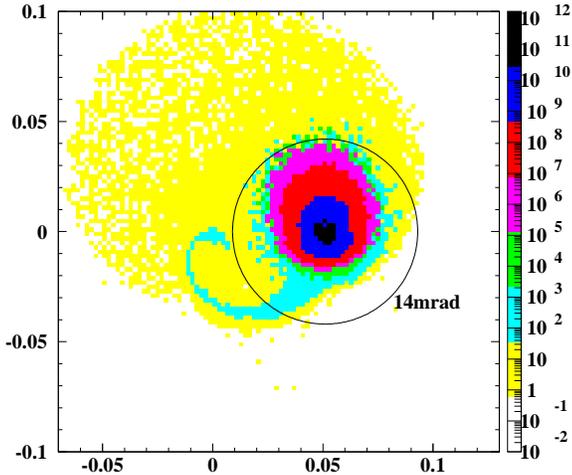}
}
\caption{The energy distribution on the calorimeter face at a distance of 3.8 m
away from the interaction point.}
\label{fig:3}       
\end{figure}

A problem at the photon collider are the high backgrounds, which can be a factor of 10 
or higher
than at the LC. Plotted in Figure \ref{fig:3} is the 
energy distribution from the beam-beam interactions at the distance of 3.8 m after the 
interaction point. The circle shows the size of the 
beam pipe of the outgoing beam. The magnetic field sweeps a significant energy out of 
the beam pipe, which amounts to more than 1 TeV/mm$^2$.

\begin{figure}
\resizebox{0.4\textwidth}{!}{%
  \includegraphics{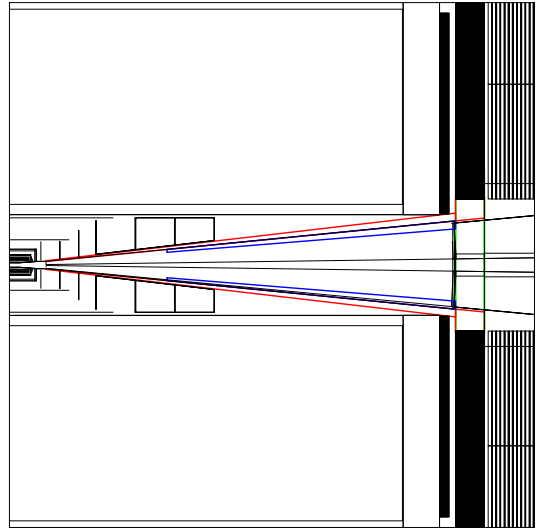}
}
\caption{A schematic view of the mask system to protect the tracking
detectors against the beam background.}
\label{fig:4}       
\end{figure}

Figure \ref{fig:4} shows the masking system design to reduce the background 
in the central tracking detectors. The whole region of the outgoing 
beams is shielded by two conical tungsten masks. 
The outer mask is extended up to the vertex detector. 
The maximum number of hits in the layers of the vertex from the incoherent pairs 
is about 372 hits, which corresponds to a hit density of 0.03 hits/mm$^2$
on the first layer. This hit density is tolerable for the current design of the 
vertex detector.
The number of photons entering the TPC is about a factor 2.4 larger than in the
$\rm e^{+} \rm e^{-}$ case, leading to an occupancy 
below 1$\%$, which can be handled in the track reconstruction. 
The detectors outside the forward region can be a copy of the LC 
detector design. 

\section{Open issues: collision stability and the beam dump}
\label{feedback}

Beam steering needs to be performed at two positions: at the photon-photon 
interaction point to keep in collision
beams with a size of 88 nm $\times$ 4.3 nm, and at the two Compton interaction points, 
with an expected laser beam size of 14 $\mu$m $\times$ 14 $\mu$m. 
Details are still to be worked out and several ideas are investigated.

At the $\gamma \gamma$ interaction point, 
first the electron beams are stabilized by a fast feedback system which measures 
the deflection of the beams, using undisrupted electron bunches at the beginning of the train. 
The steering of the photon beams relies on the fact that they follow the electron 
direction of motion and that the electron beams are stable over the entire train.
Only after the electron beams are brought into collision the laser will be turned on. 
For the Compton interaction points, it would be possible to separate electrons and 
positrons from 
the photons, and with an instrumented beam dump to measure the amount of high energy 
photons in the beam, which is an indication of a good alignment of the laser.   

The beam dump is an open subject: the problem is that the photon beam cannot be deflected
by electromagnetic fields, and this has two drawbacks. 
There is a direct line of sight from the interaction point to the dump, which is a 
problem due to the neutrons traveling back to the detector. Also, the photon beam will 
always hit the same spot on the window, causing a high thermal load and high radiation 
damage to the window. 

In conclusion, although this is a work in progress, there is no fundamental obstacle
which would stop TESLA to operate as a photon collider.
 
%
%

\section*{Acknoledgements}
I would like to thank F. Bechtel, G. Klemz, K. M\"onig, J. Sekaric and A. Stahl
for help with the preparation of this presentation.

%
%

\end{document}